\documentclass[aps,twocolumn,prb,superscript,floatfix,
superscriptaddress,showpacs,footinbib]{revtex4-2}

\usepackage{amssymb}

\usepackage{graphicx}
\usepackage{amsmath}
\usepackage{amsmath,bm}
\usepackage{blindtext, xcolor}
\usepackage{xcolor}
\usepackage{hyperref}
\hypersetup{
	colorlinks,%
	citecolor=blue,%
	linkcolor=blue,%
	urlcolor=blue
}

\usepackage{xcolor}
\usepackage{verbatim}
\usepackage{float}
\usepackage{graphicx,float}
\usepackage{silence}
\newcommand{\up}{\uparrow}

\newcommand{\uvec}[1]{\boldsymbol{\hat{\textbf{#1}}}}

\usepackage{lipsum}

\begin{document}
	
\title{Suppressed Kondo screening in two-dimensional altermagnets}%

\author{G.~S.~Diniz}
\email[e-mail:]{ginetom@gmail.com}
\affiliation{Curso de  F\'isica, ICET, Universidade Federal de Jata\'i, 
Jata\'i, GO 75801-615, Brazil.}
\author{E.~Vernek}
\email[e-mail:]{vernek@ufu.br}
\affiliation{Instituto de F\'isica,  Universidade Federal  de Uberl\^andia, Uberl\^andia, MG 38400-902, Brazil}

\affiliation{Department of Physics, Zhejiang Normal University, Jinhua 321004, Zhejiang, People’s Republic of China.}

\altaffiliation[This work was concluded while visiting the ]{Department of Physics, Zhejiang Normal University, Jinhua 321004, Zhejiang, People’s Republic of China.}
\date{\today}%

\begin{abstract}
We have studied the Kondo effect of a spin-1/2 impurity coupled to a two-dimensional altermagnet host material. To attain  the low-temperature many-body Kondo physics of the system, we have performed a numerical renormalization group calculations that allows us to access the spectral properties of the system at zero temperature. The impurity spectral function and the Kondo temperature were calculated for different set of parameters, including Rashba spin-orbit coupling (RSOC) and an external magnetic field. Interestingly, in the RSOC and altermagnetic fields, the hybridization function is spin independent, despite the characteristic broken time-reversal symmetry of the altermagnet. This is because the alternating sign of the spin splitting of the bands in the momentum space renders equal contributions for both spin components of the hybridization function. Our results demonstrate that, although the hybridization function is time-reversal symmetric, the Kondo temperature is substantially suppressed by the altermagnet coupling. Moreover, we have investigated the effect of an external magnetic field applied in the altermagnet along different directions. Interestingly, we observe an important restraining of the Kondo peak which  depends strongly on the  direction of the field. This anisotropic effect is, however, masked if strong Zeeman splitting takes place at the impurity, as it shatters the Kondo-singlet state.
\end{abstract}

\maketitle

\section{Introduction}

Screening of localized magnetic moments by itinerant  electrons in metallic 
systems is paradigmatic in many-body physics in condensed 
matter. This phenomenon, known as the Kondo effect~\cite{hewson_1993}, is 
crucial for understanding both electronic transport and magnetic properties of 
metallic systems doped with magnetic atoms. The physical mechanism underlying this phenomenon was first proposed by Kondo in 1963~\cite{10.1143/PTP.32.37}, as spin-flip scatterings of conduction electrons by magnetic impurities diluted in metallic systems, resulting in a minimal resistivity observed at low temperatures~\cite{DEHAAS1936440}. After more than fifty years of its discovery, Kondo effect has been studied in a variety of systems resulting in an abundance of scientific studies \cite{Jiang2018,Kurzmann2021,Piquard2023,PhysRevB.108.195123}.

Since the Kondo effect results from an effective exchange interaction 
between the localized moments and itinerant electrons, individual characteristics of both may be equally important. Therefore, the conditions to which the conduction electrons are subject to may modify the screening of localized magnetic moments. For example, spin-polarized (ferromagnetic) bands are detrimental to Kondo screening because time reversal symmetry (TRS) --- which is of key importance to 
spin-flip scattering process involved in the Kondo mechanism --- is broken \cite{PhysRevLett.91.127203,PhysRevLett.91.247202,science.1102068}. 
Another example is the electron-phonon interaction in the conduction band 
that results in a vanishing density of states of the conduction band at the 
Fermi level and leading to rich Kondo screening phases. More recently, 
Kondo screening of magnetic impurities by spin-orbit coupled conduction 
electrons has been investigated~\cite{Zitko2011,PhysRevB.93.075148}. The 
fundamental question in this context is how spin-orbit coupling (SOC) modifies the 
Kondo temperature of the system~\cite{PhysRevLett.108.046601}.

Quite interestingly, although SOC produces an important modification in the energy bands of the host materials, numerical renormalization analysis has shown 
that, at least in the high density regime, the effect is straightforward. The increase of SOC produces a small widening of the conduction band accompanied by a decrease of the density of states at the Fermi level. Nonetheless, the effective hybridization between the impurity and the conduction electrons around the Fermi level decreases resulting in a decreasing of $T_K$ 
\cite{PhysRevB.102.155114}. It seems, therefore, that despite the fact that 
SOC produces spin-splitting in the energy bands, because TRS is preserved, 
it has a minor effect in the Kondo screening~\cite{PhysRevLett.128.027701}. A natural follow up question one may ask is whether this is the case for any system with spin-splitting energy bands, for example, the recently discovered altermagnets~\cite{JPSJ.88.123702,PhysRevB.102.014422,PhysRevX.12.040002,PhysRevX.12.040501}, with peculiar TRS broken energy bands. 

Altermagnets are a class  of collinear two- and 
three-dimensional antiferromagnets \cite{antiferromagnet} that display a large broken spin 
degeneracy, even in the absence of 
SOC~\cite{pnas.2108924118,PhysRevX.12.011028,PhysRevX.12.040501}, and they have gained considerable attention recently. Most of these studies have focused on the peculiar properties featured by  this emergent class of magnets materials, such as large anisotropic (non-relativistic) and spin-splitting in momentum space with protected symmetry points, accompanied by a zero (net) magnetization. This anisotropic spin-splitting 
leads to the prediction of remarkable physical properties such as gapless 
superconducting states with mirage gaps~\cite{wei2023gapless}, “multi-piezo” effect in ${\rm V}_2{\rm Se}_{2}{\rm 
O}$ monolayer \cite{zhu2023multipiezo}, spin-polarized Andreev levels 
\cite{PhysRevB.108.075425}, Majorana zero modes with zero net 
magnetization \cite{ghorashi2023altermagnetic}, magnon bands with an 
alternating chirality splitting \cite{smejkal2022chiral}, pronounced thermal transport \cite{PhysRevLett.132.056701}, among many others 
exotic properties.

In this paper, we study the Kondo screening of magnetic impurities in 
altermagnets. More specifically, we aim at investigating how the unusual 
magnetic properties of these materials modify the Kondo temperature of the 
system. To this end, we propose a spin-1/2 magnetic impurity coupled to an 
altermagnet host material described formally by a traditional 
single-impurity Anderson model (SIAM). To accomplish this, we employ the 
well-known numerical renormalization group (NRG)~\cite{RevModPhys.80.395} 
that allows us to obtain the relevant low-temperature physical quantities. For instance, within NRG calculations we can obtain the local density of states LDOS at the impurity site, from which we estimate the Kondo temperature of the system. Our results shows that the intriguing band 
structure of the altermagnets has interesting consequences on the Kondo physics: (i) despite 
the spin-splitted band due to broken TRS, there is no spin splitting in the impurity spectral function. This is because the integration of the self-energy over the entire Brillouin zone renders a spin-independent hybridization function. (ii) By comparing the effects of AM and RSOC to the Kondo temperature, we find that $T_{\rm K}$ is up to two order of magnitude smaller for AM coupling as compared to RSOC, for the \emph{same strength}. 
(iii) In the presence of external magnetic field, $\bm{B}$, on the altermagnetic-magnetic impurity system, the suppression on the Kondo peak depends strongly on the direction of the applied field. This anisotropy is akin to what was obtained by one of us in quantum impurity coupled to SOC quantum wires~\cite{PhysRevB.102.155114}.

The rest of this paper is structured as follows. In Sec.~\ref{secII} we 
present the model Hamiltonian and the procedure to calculate 
the hybridization function. Sec.~\ref{secIII} the NRG results are presented. We 
summarize and conclude in Sec.~\ref{secIV}.

\section{Model and methods}
\label{secII}
\begin{figure}[h!]
\includegraphics[scale=0.6]{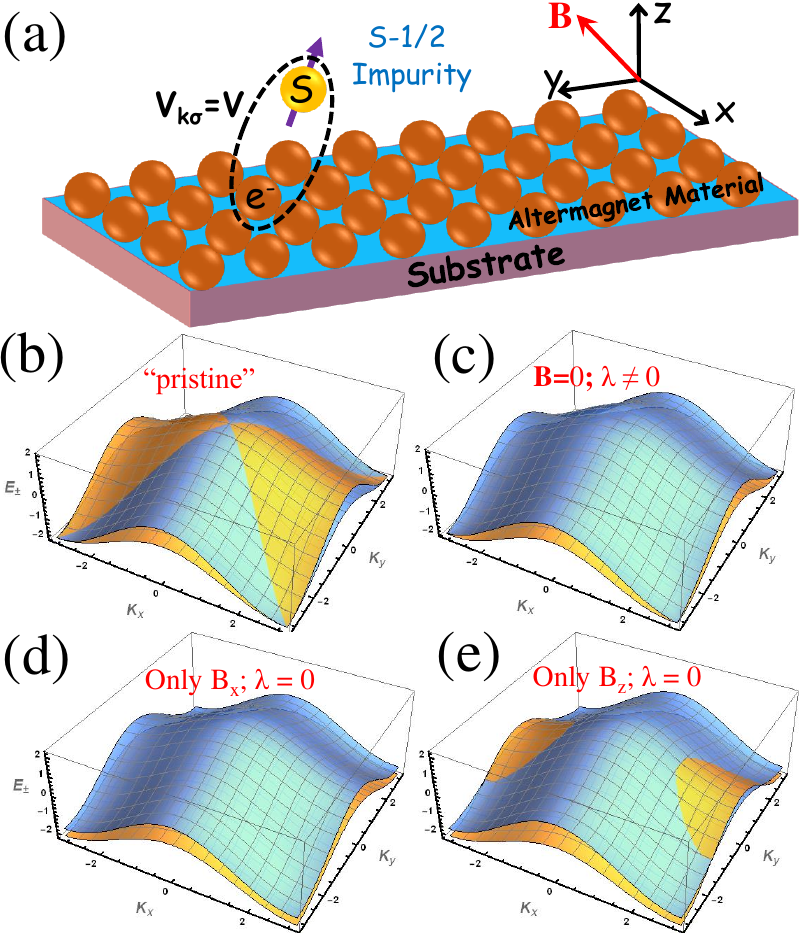} 
\caption{(a) Schematic picture showing a spin-1/2 impurity coupled to   an 
altermagnet material, which is deposited on a substrate and under an 
external magnetic field represented by the red arrow. (b) The 
$E_\pm(k_x,k_y)$ for \emph{pristine} altermagnet ($\lambda$ and ${\bm B}$ 
are both zero) the zone boundary $-\pi/a \le k_{x (y)}\le\pi/a$. (c) 
$E_\pm(k_x,k_y)$ for the case when $\lambda\neq$0 and ${\bm B}=0$. (d) 
$E_\pm(k_x,k_y)$ for the case when $\lambda$=0 and $\vec{B}=B\uvec{x}$ (or 
${\bm B}=B{\uvec{y}}$ by symmetry). (e) Same as in (d), but for a magnetic 
field along the ${\bm B}=B\uvec{z}$.
\label{fig1}}
\end{figure}

\subsection{Hamiltonian model}
We consider a hybrid system composed by a spin-$1/2$ impurity coupled to  a 
$d$-wave AM with RSOC \cite{PhysRevB.108.184505}, as illustrated in  Fig.~\ref{fig1} (a). The total 
Hamiltonian of the system can be written in the form of a SIAM as,
\begin{eqnarray}
H = H_{\rm 0} + H_{\rm imp} + H_{\rm hyb},
\label{eq:Hamilt0}
\end{eqnarray}
where $H_{\rm 0}$, $H_{\rm imp}$ and $H_{\rm hyb}$ represent the host altermagnet material, the magnetic impurity and the coupling between electrons of the altermagnet and the magnetic impurity, 
respectively. More precisely, $H_{\rm 0}= H_{\rm AM} + H_{\rm RSOC} + H_{\rm Z}$, where $H_{\rm AM}$ describes, for instance, a $d$-wave altermagnet on a substrate~\cite{PhysRevX.12.011028,PhysRevX.12.040501,PhysRevB.108.184505}. $H_{\rm RSOC}$ accounts for the 
contribution of the Rashba SOC (RSOC) induced by an inversion asymmetry, 
which can naturally arise due to the substrate on which the AM is 
deposited. Finally, $H_{\rm Z}$ represents an external magnetic field. More explicitly, each of these terms  are given  by
%
\begin{eqnarray}
\label{eq:Hamilt1}
&&  H_{\rm AM}= 2t(\cos k_{x}\!+\!\cos k_{y})\sigma_{0} \!+\!  2t_{\rm AM}(\cos 
k_{x}\!-\!\cos k_{y})\sigma_{z},\nonumber\\ 
&& H_{\rm RSOC}= 2\lambda(\sin k_{y}\sigma_{x}-\sin  
k_{x}\sigma_{y})=\bm{B}_{\rm SOC}({\bm k})\cdot{\boldsymbol\sigma},\\
&& H_{\rm Z}= g\bm{B}\cdot \bm{S}\nonumber.
\end{eqnarray}

%
Here,  $t$ is the hopping amplitude parameter and $t_{\rm AM}$ 
stands for the AM exchange interaction. We have written $H_{\rm RSOC}$ as an effective $k$-dependent \emph{magnetic} field ${\bm B}_{\rm SOC}({\bm k})=2\lambda(\sin k_{y},-\sin 
k_{x})$ induced by the Rashba field, with $\lambda$ standing for the strength of the interaction. The last term  of Eq.~\eqref{eq:Hamilt1}, $H_{Z}$,  describes 
the Zeeman effect in the AM material due to an external 
magnetic field ${\bm B}$, in which $g$ is the effective $g$-factor of 
the AM. In all these terms, $\sigma_i$ represents the   Pauli matrices, including the identity $\sigma_0$ and $\mathbf{S}=\hbar\boldsymbol{ \sigma}/2$. 

The second term of Eq.~\eqref{eq:Hamilt0} is given by
\begin{eqnarray}
H_{\rm imp}&=& \sum_{\sigma}\varepsilon_{d} n_{\sigma} +  
Un_{\uparrow}n_{\downarrow} + g_{\rm imp}\bm{B}\cdot \bm{S}, 
\label{eq:Hamilt2}
\end{eqnarray}
where, $\varepsilon_{d}$ is the impurity on-site energy for spin  $\sigma$ 
($\uparrow$ or $\downarrow$), $g_{\rm imp}$ is the impurity $g$ factor, $U$ 
represents the $e$-$e$ interaction at the impurity site, and $n_{\sigma}= 
d^\dagger_{\sigma}d_{\sigma}$ is the electron number operator with spin 
$\sigma$.

Finally, the third term of Eq.~\eqref{eq:Hamilt0} has the form
\begin{eqnarray}
H_{\rm hyb}\!=\!V\sum_{k\sigma}(d^\dag_{\sigma}c_{k\sigma} + H.c.), 
\label{eq:Hamilt3}
\end{eqnarray}
where we have assumed for simplicity, that the coupling strength  between 
the impurity and the altermagnet is independent of momentum and spin 
component~\cite{microscopic}.

\subsection{Energy band properties}
The main features of the electronic bands given by Eq.~\ref{eq:Hamilt1} are 
shown in Figs.~\ref{fig1}(b)-\ref{fig1}(e), for different sets of 
parameter configurations. Fig.~\ref{fig1}(b) shows the \emph{pristine} 
case, which shows the alternating spin-splitting along the momentum and 
with degenerate points at $\Gamma$ and also at the four corners of the 
Brillouin zone. In Fig.~\ref{fig1} (c), we show the case where $\lambda\ne$ 
0 (while ${\bm B}$=0), which fully breaks the spin degeneracy of the bands, 
except at $\Gamma$ points and  at the four corners. In 
Figs.~\ref{fig1}(d)-\ref{fig1}(e), a ${\bm B}$ is considered along different 
directions (with $\lambda$=0), where now there is no degeneracy at $\Gamma$ 
and at the boundaries of the Brillouin zone. Below, we will return to the host band structure with a deeper analysis under different interactions.

\begin{figure}[h!]
\includegraphics[scale=0.54]{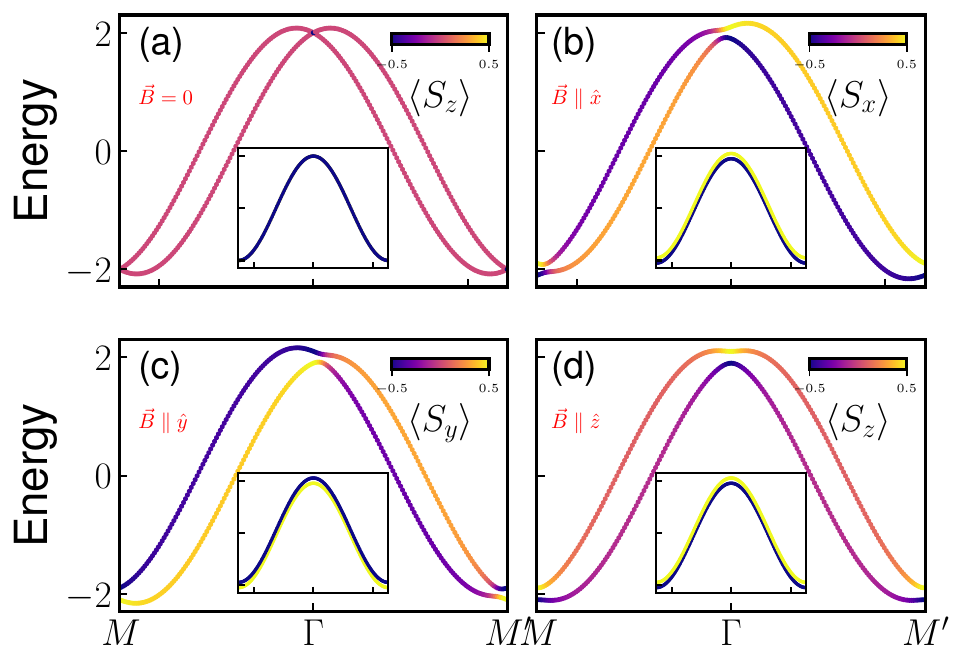} 
\caption{Energy bands along the $M$-$\Gamma$-$M^{\prime}$ path  for 
different parameters configurations: (a) ${\bm B}=0$, (b) ${\bm 
B}=0.05\uvec{x}$, (c) ${ \bm B}=0.05\uvec{y}$, and (d) ${\bm 
B}=0.05\uvec{z}$. For all panels, $t$=0.5, $t_{\rm AM}$=0.1, $\lambda$=0.2, a 
N\'eel vector along the $z$-direction and the color map is associated to 
the spin-projection, $\langle S_{i}\rangle$, for each band. The insets 
correspond to each respective cases with $\lambda$=0.0.
\label{fig2}}
\end{figure}

Let us further analyze the host band structure in the presence of the external magnetic field. For the sake of numerical convenience, throughout 
the paper we will set, $g=k_{B}=1.0=\mu_{B}=\hbar=1.0$. Fig.~\ref{fig2}(a)-\ref{fig2}(d) show  the electronic band structure 
along the path $M$-$\Gamma$-$M^{\prime}$, which is a good representative of 
the symmetries of the bands. In the absence of magnetic field and RSOC, the 
system shows an intriguing broken TRS with a momentum 
dependent alternating spin-splitting, as shown in Fig.~\ref{fig1}(b). The 
band degeneracy along the path $M$-$\Gamma$-$M^{\prime}$ is broken when 
RSOC is considered. However, they are still degenerated at the $\Gamma$ and 
$M$ ($M^{\prime}$) points, which is a consequence of the 
$C_{4z}\mathcal{T}$ symmetry that still holds~\cite{PhysRevB.108.184505}. 
This feature can be clearly seen in Fig.~\ref{fig2} (a), where we show the 
band structure for ${\bm B}$=0 and $\lambda=0.2$ ($\lambda=0.0$ for the 
inset). Moreover, the spin projection $\langle S_{z}\rangle$ is zero except 
at the spin-degenerate points, where the value is exactly $\langle 
S_{z}\rangle$=0.5 (and -0.5). Indeed, the spin projection would be non-zero (and 
equal, $\langle S_{x}\rangle$=$\langle S_{y}\rangle$) only along the plane 
($x-y$) (not shown in the panel). For an applied magnetic field in the 
plane $x$-$y$, say, at the $x$-(or $y$-direction),  
$C_{4z}\mathcal{T}$-symmetry is broken as shown in 
Fig.~\ref{fig2}(b) and \ref{fig2}(c). We observe  clear spin splittings of the bands at all 
points including $\Gamma$ and $M$ ($M^{\prime}$), which acquires a 
finite gap of $2g\vert {\bm B}\vert$, even for the $\lambda$=0 case (shown 
in the insets). However, differently from the $\lambda$=0, the presence of 
RSOC induces a spin mixing of the bands with a broken inversion symmetry 
along the $M$-$\Gamma$ and $\Gamma$-$M^{\prime}$ path. Moreover, by alternating 
the direction of the in-plane ${\bm B}$ field, the band structure around the 
$\Gamma$ point is reversed, with significant changes in the spin projection 
along the direction of the applied ${\bm B}$ for $\lambda \ne 0$, with spin 
projections inverted at the $\Gamma$ an $M$ ($M^{\prime}$) points. Lastly, 
in Fig.~\ref{fig2}(d) we show the case when $\bm{B}$ is applied along the 
$z$-direction. As previously observed, the system is TRS-broken with a clear spin splitting. However, differently from the in-plane case, the system still has a symmetry along the cut line $M$-$\Gamma$-$M$. From these results, we can confirm that the interplay between RSOC and magnetic field introduces important modifications to the altermagnet electronic bands, with a relevant anisotropy that can be crucial for the Kondo physics, as will be further explored in the following sections.

\subsection{Hybridization function}
The hybridization function of the impurity is of key importance in the NRG approach to the Kondo impurity 
problem. To obtain it, we first calculate the local impurity Green's function ${\bm G}_{imp}(\omega)$, which 
is a $2\times 2$ matrix and can be written 
as \cite{PhysRevB.92.121109,PhysRevB.102.155114},
\begin{eqnarray}
{\cal G}_{\rm imp}(\omega)\!=\!\left[(\varepsilon_{d}-\omega)\sigma_0-\boldsymbol{\Sigma}^{(0)}(\omega)-\boldsymbol{\Sigma}^{({\rm int})}(\omega)\right]^{-1}.
\label{eq:Hybrid0}
\end{eqnarray}
Here, $\boldsymbol{\Sigma}^{(0)}(\omega)=\sum_{k}\bm{V}  
\boldsymbol{G}_{\rm host}({\bm k},\omega)\bm{V}^{\dagger}$ represents the non-interacting self-energy~\cite{PhysRevB.92.121109}, 
with $\hat{V}=V\sigma_0$ and $\boldsymbol{G}_{\rm 
host}(k,\omega)=\left[\omega\sigma_0 -H_{0}\right]^{-1}$, while 
$\boldsymbol{\Sigma}^{(\rm int)}(\omega)$ is the  interacting part (or proper) self-energy, which is obtained within the NRG calculation. With the 
non-interacting self-energy, the hybridization function can be written as, 
\begin{eqnarray}
\boldsymbol{\Gamma}(\omega)\!=\!\frac{1}{2i}\! \int \! \!\!\int \left[\boldsymbol{\Sigma}^{(0)}({\bm k},\omega-i\eta)-\boldsymbol{\Sigma}^{(0)}(\bm k{},\omega+i\eta)\right]d^{2}k, 
\label{eq:Hybrid1}
\end{eqnarray}
where $\eta\rightarrow 0$. The integration defined in  
Eq.~\eqref{eq:Hybrid1} is highly peaked for small values of $\omega$, 
therefore a high precision numerical integration is desirable to avoid 
numerical flaws. To achieve this purpose, we have used cubature rules as 
implemented in the Cuba package \cite{HAHN200578}. Note that  $\boldsymbol{\Gamma}(\omega)$ is also a $2\times 2$ matrix and, in 
the presence of an external in-plane magnetic field, there is a spin-mixing 
rendering non-diagonal elements. To deal with the spin-mixing 
channels, a non-trivial Wilsonian RG chain is  
required~\cite{PhysRevB.102.155114}. As the hybridization matrix has, in general, 
non-zero off-diagonal complex elements, it is useful to decompose it in terms of Pauli 
matrices as, $\boldsymbol{\Gamma}(\omega)=\sum_{i}d_{i}(\omega)\sigma_{i}$ 
($i$=0, x, y, z). For convenience, we hereafter will rescale the bandwidth to unity ($D$=1), such 
that the relevant  energy range is defined as $-1< \omega<1$, a standard procedure in quantum impurity solvers. We also assume the impurity-altermagnet coupling to be $V=$0.1. 

\begin{figure}[h!]
\includegraphics[scale=0.54]{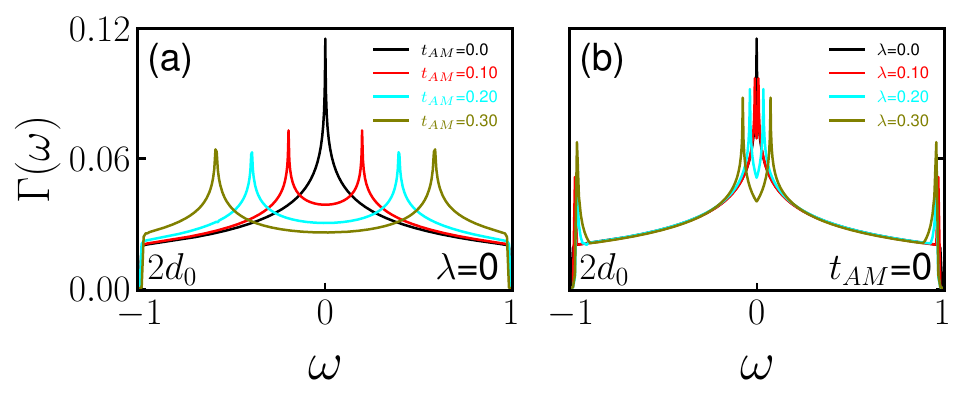} 
\caption{Hybridization function $\hat{\Gamma}(\omega)$: (a) for ${\lambda}$=0.0 and $t_{\rm AM}\neq$ 0; (b) $t_{\rm AM}$=0.0 and $\lambda\neq$0. For both panels $\bm{B}$=0, and notice 
that we are showing the trace of $d_{0}\sigma_{0}$, as the diagonal terms are equal and there are no off-diagonal terms.
\label{fig3}}
\end{figure}

In Figs.~\ref{fig3} (a) and 3(b), we show $\Gamma(\omega)$ for different values of $t_{\rm AM}$ and $\lambda$, respectively, in the absence of $\bm{B}$. Notice that regardless of the value of $\lambda$ or $t_{\rm AM}$, the diagonal terms for both panels are equal, as there is no spin-mixing terms in the hybridization matrix. Therefore, we only showed the trace of $d_{0}(\omega)\sigma_{0}\propto 2d_{0}$, which is proportional to the total density of states. When $t_{\rm AM}$ is considered (with $\lambda$=0), one can observe a strong suppression of the peak around $\omega=0$, with the development of a splitting. This suppression will be responsible to signatures in the Kondo temperature, as it will be discussed in the following section. For the case when RSOC is non-zero (with $t_{\rm AM}$=0), shown in Fig.~\ref{fig3} (b),
we can also notice a peak suppression similar to the Fig.~\ref{fig3} (a), however, the suppression is less pronounced, and there are peaks at the edge associated to the van Hove singularities. Since they are far away from the Fermi level, these van Hove singularities at the edges near the edge of the conduction band are unimportant for the Kondo screening. The more pronounced suppression of the hybridization function near the Fermi level produced by $t_{\rm AM}$ anticipates the stronger suppression of Kondo screening caused by the altermagnet as compared to RSOC. 

\begin{figure}[h!]
\includegraphics[scale=0.54]{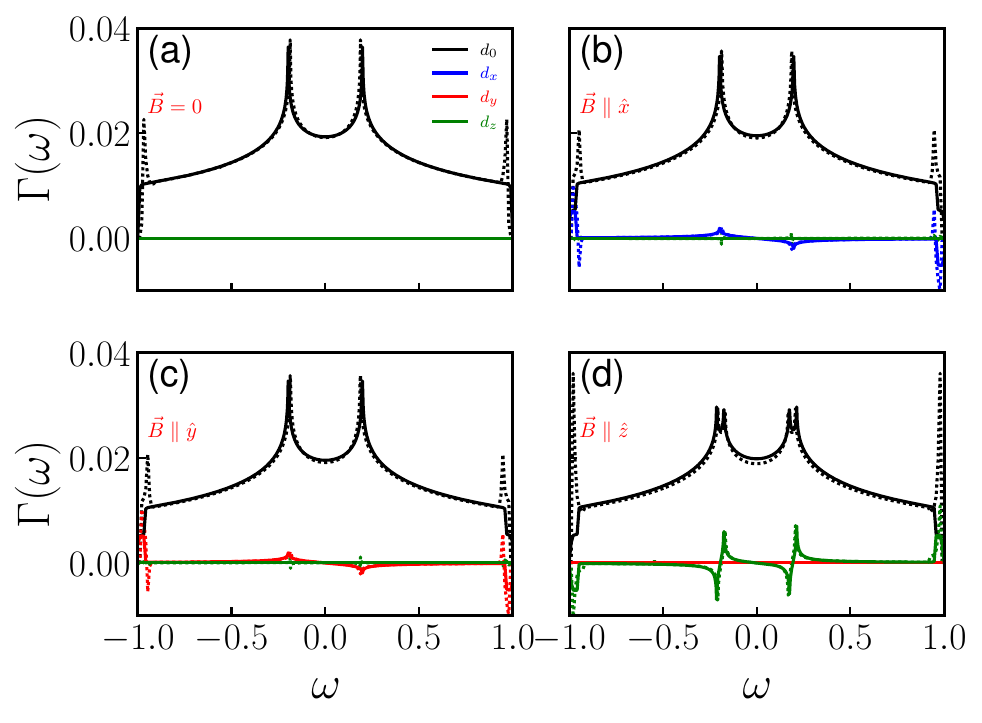} 
\caption{Decomposition of the matrix hybridization  function 
$\hat{\Gamma}(\omega)$ along different Pauli matrices $\sigma_i$ with 
corresponding weights $d_{i}$. (a) For ${\bm B}$=0, (b) for a magnetic 
field along $\uvec{x}$, (c) for a magnetic field along $\uvec{y}$ and (d) 
for a magnetic field along $\uvec{z}$. The solid line is for $\lambda$=0, 
while the dashed line stands for $\lambda$=0.2. For (b)-(d), we have 
assumed $t_{\rm AM}=0.1$ and $B=0.02$.
\label{fig4}}
\end{figure}

Now we set the value of $t_{\rm AM}$=0.1 and investigate how the presence of $\bm{B}$ affects the hybridization matrix. In Fig.~\ref{fig4}, we show $\boldsymbol{\Gamma}(\omega)$ decomposed for different parameter configurations of ${\bm B}$ and $\lambda$. In the 
absence of a magnetic field and RSOC ($\lambda$=0), $d_{0}$ is the only 
nonzero value, as shown by the black solid line in Fig. \ref{fig4} (a). 
However, notice that this result holds even for the case of $\lambda$=0.2 
(dashed line). This behavior is because (i) the impurity 
is spin-independently coupled to the host and (ii) the AM Hamiltonian [$H_0$ in 
Eq.~\eqref{eq:Hamilt0}] preserves $C_{4z}\mathcal{T}$ symmetry and  SOC is 
time-reversal invariant, thus rendering a diagonal and spin-independent 
$\boldsymbol{\Gamma}(\omega)$. Nonetheless, notice that 
because $\lambda \ne$0, we can still see the emergent peaks close to the edges of the 
conduction band associated to the appearance of van Hove singularities in the 
conduction band around the $M$-point. For an in-plane magnetic field Fig. \ref{fig4} (b) and 4(c), the off-diagonal elements of 
$\hat{\Gamma}(\omega)$ will contribute to non-zero values for the 
coefficients $d_{y}$ ($d_{x}$) for a magnetic field applied along $x$ ($y$) 
conversely, while there is also a finite value for $d_{z}$ that changes 
signs, i.e., $d_{z} \rightarrow -d_{z}$. Interestingly, for $\bm{B}\parallel 
\uvec{z}$ shown in Fig.~\ref{fig4} (d), the only nonzero components are 
along $d_{0}$ and $d_{z}$, even when $\lambda \ne$0 (which intertwine the 
spin channels but does not break TRS). Furthermore, notice that because of 
the interplay between the AM exchange interaction and the Zeeman field, 
there is an enhancement of the $d_{z}$ component, and also a marked \emph{double}
splitting in the peaks of $d_{0}$, which will strongly reflect in the impurity spectral function.

\section{Interacting regime and Kondo screening}
\label{secIII}

Having obtained the  hybridization function $\Gamma(\omega)$, we are ready to use the NRG method to address the Kondo effect in the system. This numerical technique is a powerful and standard impurity solver, which 
consists of a logarithmic discretization of the continuum conduction band 
in energy scales that decreases a $\Lambda^{-N/2}$, where $\Lambda$ is a discretization parameter and $N$ is the number of iterations in the NRG procedure. The discretized  model is then mapped into a one-dimensional tight-binding-like  Hamiltonian (aka Wilson chain Hamiltonian) whose length increase iteratively with $N$. The iterative diagonalization of 
a large-$N$ Wilson chain is possible by proper truncation of Hilbert space at certain maximum dimension, allowing us to access the energy around the Fermi level in a controllable manner~\cite{RevModPhys.47.773}. Here, for practical purposes, we employ the NRG  method to calculate the impurity retarded Green's function, from which we calculate the impurity density of states in the interacting regime. Within the Zubarev's notation~\cite{Zubarev1960}, our Green's function can be written as
\begin{eqnarray}\label{retarded_GF}
{\cal G}^{\sigma \sigma^\prime}_{\rm imp}(\omega)=\langle\langle d_\sigma;d^\dagger_{\sigma^\prime} \rangle\rangle_\omega,
\end{eqnarray}
where $\langle\langle d_\sigma;d^\dagger_{\sigma^\prime} \rangle\rangle_\omega=-i\int \Theta(t)\langle \left[d_\sigma(t),d_{\sigma^\prime}^\dagger(0) \right]_+ \rangle e^{i\omega t}dt$. Here, $\left[d_\sigma(t),d_{\sigma^\prime}^\dagger(0) \right]_+$ is the anti-commutator between the 
annihilation operator at time $t$ with the creator operator at $t=0$ at the impurity. The 
expectation value $\langle \cdot \rangle$ is taken in the ground state for zero temperature calculations.

The nontrivial energy dependence of the hybridization function requires an improved discretization scheme to reproduce with high resolution the initial 
function, thus reducing the numeric artifacts that can be present in the NRG 
calculations. To this end, we have used the adaptive 
$z$-averaging scheme \cite{PhysRevB.79.085106} (with $N_{z}$=5), as 
implemented in the NRG Ljubljana open source code 
\cite{zitko_rok_2021_4841076}. Furthermore, for the spectral functions 
calculations we have also checked with the self-energy trick, as described in 
Ref. \cite{RBulla_1998} to analyze and remove nonphysical oscillations at the vicinity 
$\omega = 0$. For all the NRG calculations, we have assumed 
(unless stated otherwise) $\varepsilon_{d}=-U/2$ (particle-hole symmetry), 
U=0.25, $\Lambda$=2, and number of kept states = 2000.

\subsection{Kondo temperature}	
We start by investigating the effect of the AM exchange field in the Kondo temperature of the system. The main quantity calculated here is the impurity density of states \cite{spindiagonal}: 
\begin{eqnarray}
A_\sigma(\omega)=-\frac{1}{\pi}{\rm Im}[{\cal G}^{\sigma\sigma}_{\rm imp}(\omega)].
\end{eqnarray}

\begin{figure}[h!]
\includegraphics[scale=0.54]{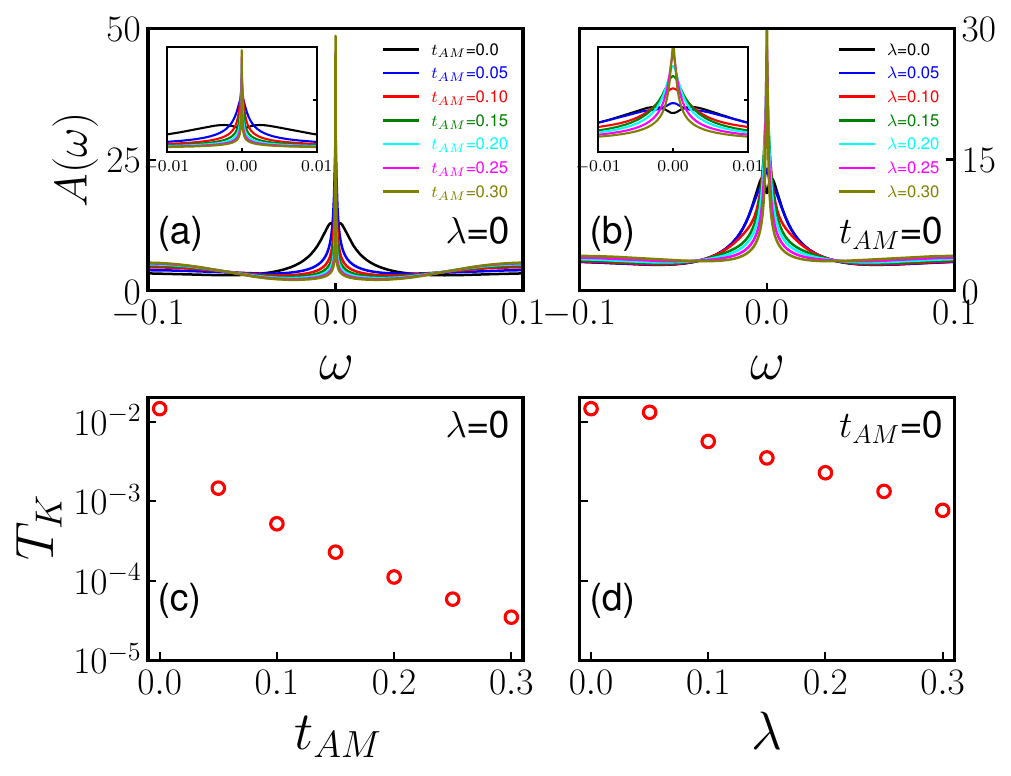} 
\caption{(a) Impurity spectral function $A(\omega)$ for  different values 
of $t_{\rm AM}$ with $\lambda$=0. (b) Impurity spectral function $A(\omega)$ 
for different values of $\lambda$ with $t_{\rm AM}$=0. (c) $T_{K}$ as function 
of $t_{\rm AM}$. (d) $T_{K}$ as function of $\lambda$. For all panels, $\bm{B}$ is set 
to zero. The insets in Fig. \ref{fig5} (a)-(b) are zooms for the range of $\omega$ in [-0.01,0.01] for better visualization.
\label{fig5}}
\end{figure}

It is  known that $A(\omega)=\sum_\sigma A_\sigma(\omega)$ exhibits a many-body resonance at the Fermi level, known as Abrikosov-Shul resonance~\cite{Wiegmann1980,Tsvelick1983} (or simply Kondo peak), as a signature of the Kondo screening~\cite{hewson_1993}. This resonance not only signals the presence of  Kondo screening in the system, but also provides a useful way to estimate $T_K$. Indeed, the width of the Kondo resonance can be directly associated to $T_K$~\cite{Nozires1974,PhysRevLett.70.4007,PhysRevB.108.L161109}. In Fig.~\ref{fig5}(a)-\ref{fig5}(d), we make an insightful analysis on the  behavior of $T_{K}$, extracted from the half-width at half-peak of the impurity spectral function, for different configurations of $t_{\rm AM}$ and $\lambda$, setting ${\bm B}=0$. Fig.~\ref{fig5}(a) shows  $A(\omega)$  for different values of $t_{\rm AM}$. For $t_{\rm AM} = 0$ (black line), we see  that $A(\omega)$ exhibits a split peak around $\omega=0$. This splitting is induced by the sharp peak in the hybridization function $\Gamma(\omega)$ shown in Fig.~\ref{fig3}(a) for $t_{\rm AM}=0$ (black curve). This splitting is analogous to the one observed in Ref.~\cite{PhysRevLett.97.096603}, resulting from an Anderson impurity coupled to a structured host density of states provided by a non-interacting energy level. Here, on the other hand, the sharp structure  the conduction band is naturally provided by the van Hove singularity of the two-dimensional material. Now, as $t_{\rm AM}$ increases, we observe that the Kondo peak evolves rapidly to a single and sharper peak, as already observed for $t_{\rm AM}=0.05$. This is  because the sharp van Hove singularity is split for finite $T_{\rm AM}$, leaving $\Gamma(\omega)$ relatively smooth near the Fermi level, as  seen in  Fig.~\ref{fig3}(a) for finite $t_{\rm AM}$. This prompt sharpening of the Kondo resonance, as $t_{\rm AM}$ increases reveals a strong dependence of $T_{\rm K}$ with $t_{\rm AM}$. Fig.~\ref{fig5}(c) shows how $T_{\rm K}$ decreases as $t_{\rm AM}$ increases as anticipated earlier in this paper. 

To compare the effects of $t_{\rm AM}$ to $\lambda$ on $T_{\rm K}$, in Fig.~\ref{fig5}(b) shows how $A(\omega)$ is affected by the SOC $\lambda$. We now set  $t_{\rm AM}=0$. We note that $\lambda$ produces, qualitatively, a very similar effect of $t_{\rm AM}$, although much less pronounced, as $\lambda$ produces a very modest splitting of the $\Delta(\omega)$. Indeed, the effect of  $t_{\rm AM}$ on $T_{\rm K}$ is much more important as compared to the effect of $\lambda$. By comparing Figs.~\ref{fig5}(c) and 5(d), we observe that $t_{\rm AM}=0.3$ produces a suppression of $T_{\rm K}$ by almost three orders of magnitude, while for the same value of $\lambda$ the suppression is about one order of magnitude. 
\begin{figure*}[thb!]
\centering
\includegraphics[scale=1.12]{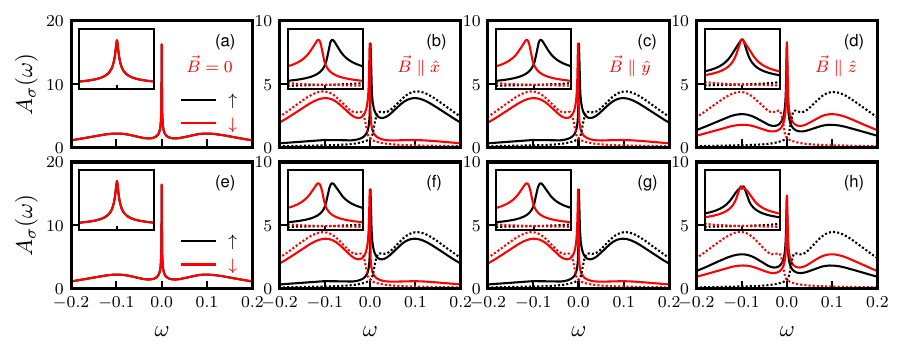} 
\caption{Impurity spin-resolved local density of states  
$A_{\sigma}(\omega)$ projected along the direction of applied magnetic 
field $\mathbf{B}$. (a)-(d) for $\lambda=0$ and $\bm{B}=0, \, 
0.002\uvec{x},\, 0.002\uvec{y}, \,0.002\uvec{z}$, respectively. (e)-(h) are 
the same results, but for $\lambda=0.2$. Solid lines (dashed) are for 
$g_{imp}$=0.0 ($g/g_{imp}$=2.5) and $t_{AM}$=0.1 for all panels. The insets are zooms for the respective 
panel in the $\omega$ range [-0.005, 0.005].   
\label{fig6}}
\end{figure*}

Here, we should mention that the dependence of $T_{\rm K}$ with $\lambda$ has been a subject of interest over the past decades, and it is still source of debate in the literature \cite{PhysRevLett.128.027701}. For instance, Zarea \textit{et al.} \cite{PhysRevLett.108.046601} reported that away from the particle-hole symmetry in a two-dimensional electron gas (2DEG), the parity breaking Dzyaloshinsky-Moriya term is able to renormalize the antiferromagnetic Kondo coupling with an exponential enhancement of $T_{K}$ \cite{PhysRevLett.108.046601}. A similar result has also been found by one of us for a quantum wire, even at the particle-hole symmetric point \cite{PhysRevB.94.125115}. Interestingly, depending on how drastic the modification of conduction band is, we can also 
observe a suppression of $T_{K}$, as the RSOC is increased, as reported for a quantum wire in the absence of $\bm{B}$ \cite{PhysRevB.102.155114}. 

Within the NRG approach, its is quite clear that the change in the Kondo temperature is directly associated to the modification around the Fermi level in the density of states of the host material induced by properties of the band structure. Therefore, for a given fixed coupling $V$ as in Eq.~\eqref{eq:Hamilt3}, enhanced density of the host material produces an enhancement of the $T_{\rm K}$. A splitting of the Kondo peak is produced if the host density of state, $\rho(\omega)$ exhibit a peak at $\omega=0$ whose width is similar or smaller than $T_K$. See, for instance, the discussion in Ref.~\cite{PhysRevLett.97.096603}.

\subsection{Effect of magnetic field}	
In Sec.~\ref{secII}, we have seen that the effect of magnetic field applied in the altermagnet is anisotropic. Therefore, anisotropic suppression of the Kondo should be observed. To see this, in Fig.~\ref{fig6} we show the spin-resolved local density  of states (LDOS) at the impurity, $A_{\sigma}(\omega)$, for the magnetic field applied (${B}=0.02$) along different directions. We have set the $g$-factor ($g=1$) for the host material (the altermagnet). The first and second rows show, respectively, results for $\lambda=0.0$ 
and $\lambda =0.2$ while solid and dashed lines corresponds to $g_{\rm imp}=0$ and $g_{\rm imp}=0.4 g$, respectively.

Let us start looking at Fig.~\ref{fig6} (a) where there is no magnetic field 
applied, therefore all the projected components of $A_{\sigma}$ are equal 
to each other. Interestingly, even though TRS is broken by AM exchange, because of the unusual spin splitting in momentum space, both spin-components of the $\bm{G}_{\rm host}$ contributes 
equally to the integral of Eq.~(\ref{eq:Hybrid1}). As expected, an inversion symmetry breaking induced by RSOC does not split 
$A_{\sigma}(\omega)$, as shown in Fig.~\ref{fig6} (e). Moreover, the results are almost unaffected when  RSOC is included. Indeed, as observed in the insets of Figs.~\ref{fig6}(a) and \ref{fig6}(b), the major splitting in the hybridization function is produced by $t_{\rm AM}$.

Now, intriguing behavior occur when the magnetic field is considered. Figures ~\ref{fig6}(b)-6(d) and \ref{fig6}(f)-\ref{fig6}(h) show LDOS projected along the  different directions of applied $\bm{B}$. As such, for instance, if the magnetic field is applied along the $x$-direction, $\up$ and $\downarrow$ refer to spin pointing towards positive and negative $x$-direction \cite{projection}.
Keeping this in mind, we observe, that regardless of the direction of the applied $\bm{B}$, a suppression of the Kondo peak is accompanied by  splitting of the Kondo peak. However, for an in-plane magnetic field, its interplay with the AM term and with  $\bm{B}_{\rm SOC}$  brings a remarkable anisotropic effect in the LDOS as shown in Fig.~\ref{fig6}(b) and \ref{fig6}(c). Indeed, we see a stronger 
suppression of $A_{\sigma}(\omega)$ as compared to $\bm{B}$ along the 
$z$-direction in Fig.~\ref{fig6}(d) and 6(h). This anisotropic effect is observed only if the direct Zeeman effect is negligible in the impurity ($g_{\rm imp}\approx 0$). If we consider $g_{\rm imp}=0.4$ as shown in dashed lines, the Kondo resonance peak is almost fully suppressed, as at this magnetic field strength both spins (band and impurity) are fully polarized, thus completely destroying the Kondo singlet state.

\begin{figure}
\includegraphics[scale=0.54]{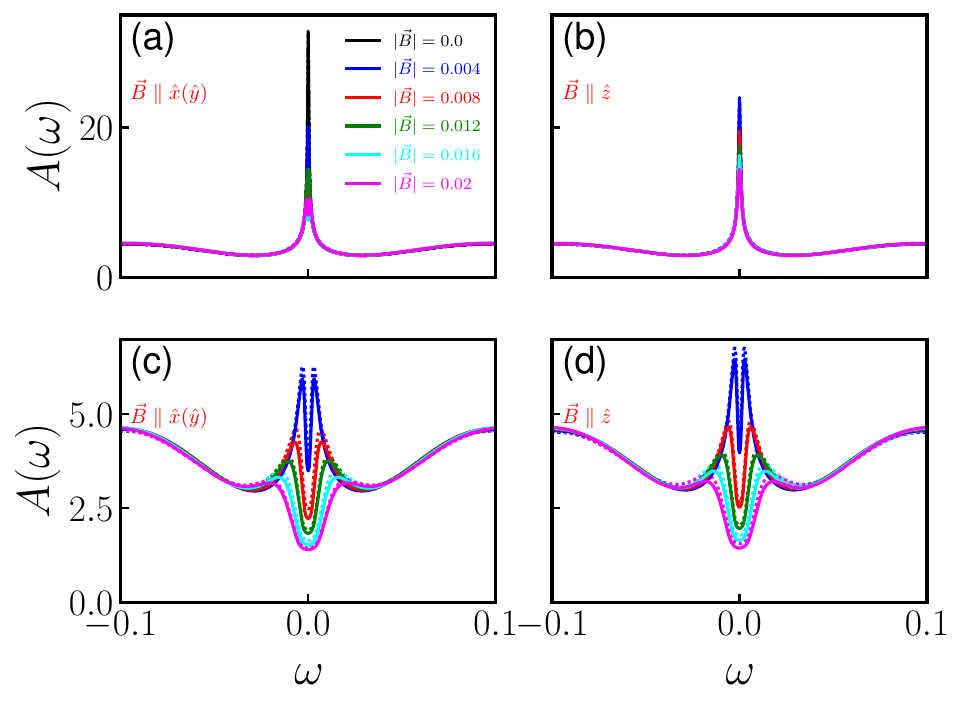} 
\caption{Impurity local density of states $A(\omega)=\sum_{\sigma} 
A_{\sigma}(\omega)$ projected along the direction of applied magnetic field 
$\mathbf{B}$. (a) $A(\omega)$ along $\uvec{x}$ (or $\uvec{y}$, by symmetry) 
for $g_{\rm imp}=0$ and $B=0, \,0.004,\, 0.008,\, 0.012,\, 0.016$, and $0.020$, 
respectively. (b) Same as in (a), but  for $A(\omega)$ along 
$\uvec{z}$. (c), (d) are for $g/g_{\rm imp}=2.5$. Solid lines (dashed) are 
for $\lambda$=0.0 ($\lambda=0.2$) and $t_{AM}$=0.1 for all panels.
\label{fig7}}
\end{figure}

Before concluding, in Fig.~\ref{fig7} we show how the anisotropic magnetic response manifests in the total density of states, $A(\omega)$. To this end, we show $A(\omega)$  for different strengths and direction of the applied magnetic field. Upper and lower panels correspond to $g_{\rm imp}=0$ and $g_{\rm imp}=0.4g$. For $B=0$, the Kondo peak can be clearly seen as shown by the  black line in Fig~\ref{fig7}(a). Now, as $\bm{B}$ increases the Kondo peak is progressively suppressed for magnetic field along either the $\uvec{x}$, $\uvec{y}$ or $\uvec{z}$ direction, as shown in Fig.~\ref{fig7}(a) and \ref{fig7}(b) for $g_{\rm imp}$=0. However, for the field along $z$-direction the peak is less affected, as compared to the other directions. Again, this is a signature of the anisotropic magnetic response of the system induced by the interplay between $\bm{B}$, AM, and RSOC. Finally, for $g_{\rm imp}=0.4g$, the Kondo peak is strongly suppressed with an enhanced splitting, as shown in Fig.~\ref{fig7} (c) and 7(d), providing support that the system no longer sustains the Kondo single state.

\section{Conclusions}
\label{secIV}
In summary, we have studied the Kondo screening effect in a spin-1/2 impurity coupled to a two-dimensional altermagnet in the presence of external magnetic field. By describing the system within a single-impurity Anderson Hamiltonian, we determined the hybridization function by exact calculation of the local Green's function, which allows us to garner the complete influence of the host material onto the impurity. We employ a NRG approach to calculate the spin-resolved  local density of states in the interacting regime, which allowed us to attain the low-temperature  Kondo physics in the system. Our results shows a drastic decrease of the Kondo temperature  $T_{\rm K}$ as the altermagnet coupling  $t_{\rm AM}$  increases. By comparing this suppression with the one produced by the RSOC $\lambda$, we showed that $t_{\rm AM}$ has a dominant effect in the suppression of $T_{\rm K}$. Furthermore, despite the TRS broken by $t_{\rm AM}$, it does not produce spin-splitting in the hybridization function. This is a direct consequence of the alternating momentum-dependent band spin-splitting that renders zero net magnetization to the impurity upon integration of the self-energy over the entire Brillouin zone. Moreover, we have investigated the effect of an applied magnetic field along different directions. The results showed an important anisotropic response of the system to the applied field. Indeed, a in-plane field  has a strong effect on Kondo screening marked by a sizable suppression of the Kondo peak. In contrast,  this suppression is less pronounced for a magnetic field along the $z$-direction. When the Zeeman coupling  takes place directly in the impurity, the Kondo peak is overwhelmingly suppressed  by the field and the anisotropic effect blurred. Our paper shed light on the Kondo correlation in altermagnets and paves the way for future theoretical as well as experimental investigations of correlated phenomena in these materials.


\acknowledgments
The authors acknowledge financial support from CAPES,
FAPEMIG and CNPq. EV thanks FAPEMIG (Process No. PPM-
00631-17) and CNPq (Process No. 311366/2021-0). 
This paper used resources of the ``Centro Nacional de Processamento de Alto Desempenho em São Paulo (CENAPAD-SP).''
%
\bibliography{references}

\begin{thebibliography}{48}
\bibitem [{mic()}]{microscopic}%
  \BibitemOpen
  \href@noop {} {\bibinfo {title} {This approximation can capture the essential
  physics of the system. however, a deeper analysis considering different
  orbitals and spin-dependent coupling between the impurity and the host might
  be relevant, specially for the microscopic description.}}\BibitemShut {Stop}%
\bibitem [{spi()}]{spindiagonal}%
  \BibitemOpen
  \href@noop {} {\bibinfo {title} {Notice that the impurity green's function is
  a full 2$\times$2 matrix, however, with non-vanishing off-diagonal elements
  only when the magnetic field $\mathbf{B}$ is applied along the $x$-$y$ plane.
  in any case, we are interested in the spin-resolved impurity spectral
  function $a_{\sigma}(\omega)$.}}\BibitemShut {Stop}%
\bibitem [{pro()}]{projection}%
  \BibitemOpen
  \href@noop {} {\bibinfo {title} {To obtain the ldos projected along different
  spin directions, we calculate the corresponding green's function, which can
  be obtained by combining appropriately the components given by
  eq.~\eqref{retarded_GF}.}}\BibitemShut {Stop}%
\end{thebibliography}
\end{document}